\documentclass[superscriptaddress,twocolumn,showkeys]{revtex4}
\usepackage{textcomp}
\usepackage{multirow}
\usepackage{amsmath}
\usepackage{amssymb}
\usepackage{graphicx}
\usepackage{epstopdf}
\usepackage{float}
\usepackage{booktabs}
\usepackage{dcolumn}
\usepackage{bm}

\usepackage{color}

\begin{document}

\title{Classifying the Cosmic-Ray Proton and Light Groups on the LHAASO-KM2A Experiment with the Graph Neural Network}

\author{Chao Jin}
\email{jinchao@ihep.ac.cn}
\affiliation{Key Laboratory of Particle Astrophysics, Institute of High Energy Physics, Chinese Academy of Sciences,Beijing 100049,China}
\affiliation{University of Chinese Academy of Sciences, 19 A Yuquan Rd, Shijingshan District, Beijing 100049, P.R.China}
\author{Song-zhan Chen}
\affiliation{Key Laboratory of Particle Astrophysics, Institute of High Energy Physics, Chinese Academy of Sciences,Beijing 100049,China}
\author{Hui-Hai He}
\affiliation{Key Laboratory of Particle Astrophysics, Institute of High Energy Physics, Chinese Academy of Sciences,Beijing 100049,China}
\author{for the LHAASO Collaboration}

\begin{abstract}
Precise measurement about the cosmic-ray (CR) component knee is essential for revealing the mistery of CR's acceleration and propagation mechanism, as well as exploring the new physics. However, classification about the CR components is a tough task especially for the groups with the atomic number close to each other. Realizing that the deep learning has achieved a remarkable breakthrough in many fields, we seek for leveraging this technology to improve the classification performance about the CR Proton and Light groups on the LHAASO-KM2A experiment. In this work, we propose a fused Graph Neural Network model in combination of the KM2A arrays, in which the activated detectors are structured into graphs. We find that the signal and background can be effectively discriminated in this model, and its performance outperforms both the traditional physics-based method and the CNN-based model across the whole energy range.
\end{abstract}
\maketitle


\section{Introduction}
Benefit from the rapid development of the computational resources, i.e. GPU, the deep learning has achieved a remarkable progress in many fields, such as the object detection and classification \cite{alexnet, yolo, faster_rcnn}, machine translation \cite{machine_translation, google_translation} and speech recognition \cite{speech_recognition, deep_speech1, deep_speech2}. While traditional methods often resolve those issues through some handcrafted features based on the expertise knowledge, the deep learning methods learn the internal representation through an end-to-end training paradigm, i.e. the Convolutional Neural Networks (CNNs) \cite{lecun_cnn, resnet}, the Recurrent Neural Networks (RNNs) \cite{rnn, lstm, gru}. The characters of the sparse connectivity and parameter sharing construct the CNN as a powerful engine in analyzing the image data, while the internal units with loops and states make the RNN efficient in modeling the time-dependent series.
\par
In essence, the success of those deep learning methods is partially owing to the effectiveness in extracting the latent representation from the regular Euclidean data (i.e. image, text, speech). Nowadays, there is an increasing number of demands for effectively analyzing the non-Euclidean data with irregular and complex structures. Proposed methods construct them as graph-structured data and exploit the deep learning to learn their representation. For instance, in e-commerce and social media platforms, the graph-based learning system exploits the interactions between the users and products to make highly accurate recommendations \cite{matrix_completion, geometric_matrix_completion}. In chemistry, the molecules are modeled as the graphs to explore and identify their chemical properties \cite{mpnn}. In the high-energy physics field, researchers need to analyze large amount of irregular signals. Some researches explore to improve their analysis efficiency with the graph neural networks (GNNs). And impressive progresses have been achieved, include improving the neutrino detecting efficiency on the IceCube \cite{icecube_gnn}, exploring SUSY particles \cite{lhc_stop} and recognizing the jet pileup structures \cite{lhc_pileup} on the LHC.
\par
Precise measurement about the cosmic-ray (CR) spectrum and their components at PeV scale is essential for probing the CRs' origin, acceleration and propagation mechanisms, as well as exploring the new physics.  A spectral break at $\sim$ 4 PeV called the CR's knee was found over 60 years \cite{cr_knee} but its origin remains a mystery. Precise localization of the component knee is the key issue for exploring the hidden physics. Current explanations of the CR knee can be classified into tow categories with different mechanisms, including the mass-dependent knee and rigidity-dependent knee models \cite{lhaaso_knee}, where models with the rigidity-dependent knee are often considered originating from the acceleration limit and the galactic leakage mechanism, and many of the models with the mass-dependent knee are associated with the new physics. Although much efforts have been paid aiming at resolving this issue, those experimental measurements still make large discrepancy with each other \cite{argo_knee, asg_knee, kascade_knee, kascade_heavy}.
\par
The Large High Altitude Air Shower Observatory (LHAASO) is the next generation of the CR experiment \cite{lhaaso_overview}. It aims at precisely measuring the CR spectrum together with their groups from 10 TeV to EeV, and surveying the northern hemisphere to identify the gamma-ray sources with a high sensitivity of 1\% Crab unity. It will be located at high altitude (4410 m a.s.l.) in the Daocheng site, Sichuan Province, China. It consists of an EAS array KM2A covering 1.3 $km^2$ area, 78000 $m^2$ closed packed water Cherenkov detector array (WCDA), and 12 wide-field Cherenkov/fluorescence telescopes (WFCTA). The LHAASO-KM2A occupies the major area and is composed by two sub-arrays, including 1 $km^2$ array of 5195 electromagnetic particle detectors (ED) and the overlapping 1 $km^2$ array of 1171 underground water Cherenkov ranks for muon detectors (MD). The WCDA contains three water ponds with the effective depth about 4 m. Each pond is divided by 5 m $\times$ 5 m cells with an 8-inch PMT located at the bottom center to watch the Cherenkov light generated by the EAS secondary particles in the water. And the focal plane camera in each telescope of WFCTA has 32 $\times$ 32 pixels with every pixel size $0.5^{\circ} \ \times \ 0.5^{\circ}$.
\begin{figure}
\includegraphics[width=.45\textwidth]{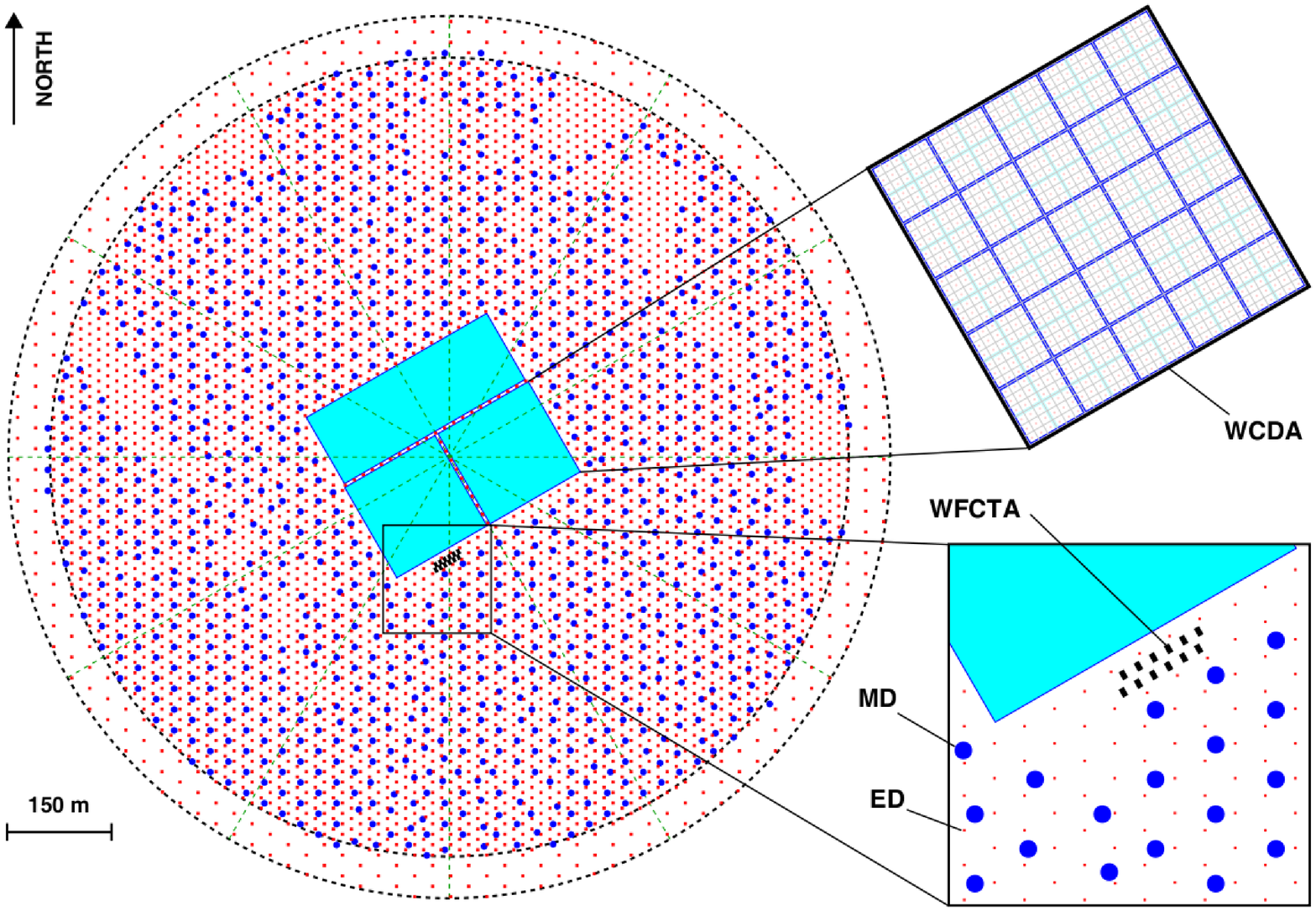}
\caption{\label{fig:lhaaso} Layout of the LHAASO experiment. The insets show the details of one pond of the WCDA, and the EDs (red points) and MDs (blue points) of the KM2A. the WFCTA, located at the edge of the WCDA, is also shown}
\end{figure}
\par
The layout of each component of LHAASO is illustrated in the Fig. \ref{fig:lhaaso}, where the red and blue points represent the KM2A-ED and KM2A-MD detectors, respectively. The ED detectors are divided into two parts, the central part with 4901 detectors and an out-skirt ring with 294 detectors to discriminate the showers with the core located within the central area from the outside ones. An ED unit consists of 4 plastic scintillation tiles (100 cm $\times$ 25 cm $\times$ 1 cm each) covered by 5-mm-thick lead plates to absorb the low-energy charged particles in showers and convert the shower photons into electron-positron pairs. The MD array plays the key role in discriminating the gamma-rays from the CR nuclei background, as well as offers the important information for classifying the CR groups. An MD unit has an area of 36 $m^2$, buried by the overburden soil with 2.5 meters high for shielding the electromagnetic components in showers. It is designed as a Cherenkov detector underneath soil, to collect the Cherenkov light induced by muon parts when they penetrate the water tank.
\par
There have been some researches on the component discrimination for the LHAASO hybrid detection with both the expertise features \cite{lhaaso_hand} and the machine learning mathods \cite{lhaaso_gbdt}. Those hybrid detection methods utilize the effective information offered by the whole LHAASO arrays. Although they exhibit a remarkable performance, their statistics are limited due to the poor operation time and aperture. Under the merit of the large area, full duty cycle, and excellent $\gamma / P$ discrimination ability, the LHAASO-KM2A is an ideal candidate for studying the CR component classification task. In this work, we leverage the GNN to improve the CR-component classification performance on the LHAASO-KM2A experiment, where the detector activated by the event is formed as the graph-structured data. Our previous work \cite{lhaaso_knee} manifested that the science issue requires high accuracy in classifying the CR Proton (P task) and Light-component (L task) from the background. Hence, we focus on these two tasks in this work. In order to evaluate the GNN performance, we also introduce the traditional physics-based method with the handcrafted feature as the baseline. The following contents are organized as follows. First, we introduce the physics baseline method in Section II. Then, we review the development of the GNN framework and propose our KM2A GNN framework in Section III. We perform the experiment and evaluate the results in Section IV and Section V. In the last Section, we make a conclusion about the GNN performance.

\section{Physics Baseline}
Current experiments detect the high-energy CRs all through the indirect methods, which measure the secondary particles of the extensive air showers (EAS) induced by the primary CR nuclei. As the CR nuclei impinging on top of the atmosphere, they suffer the hadronic interaction with the air molecules and generate daughter particles recursively, which is called the hadron cascade. The sequence of this interaction follows by the following reaction and decay schemes \cite{eas}
\begin{align*}
p + p  & \rightarrow   N + N + n_1 \pi^{\pm} + n_2 \pi^0 \\
\pi^0 & \rightarrow  2 \gamma \\
\pi^{+} & \rightarrow  \mu^{\pm} + \nu_{\mu} \\
\mu^{+} & \rightarrow  e^{\pm} + \nu_{\mu} + \nu_{e}
\end{align*}
Where the photons, electrons and positrons forms the major electromagnetic pars of the EAS and in turn generate themselves through the pair-production $\gamma \rightarrow e^+ + e^-$ and bremsstrahlung process $e^{\pm} \rightarrow e^{\pm} + \gamma$, which is called the electromagnetic cascade. Neutrinos form the missing part of the EAS and generally is ignored in experiments. The muons, without a cascade themselves, have relatively long life (2.2 $\mu$s) and  comparatively small energy loss in the media, so a large fraction of muons produced in a shower will penetrate the atmosphere and accumulated until their arrival at the observation level.
\par
The task for classifying the CR primary groups relies on the electromagnetic and muon parts of the EAS. In the first-order approximation \cite{eas_1st}, a primary CR nuclei with mass A and energy E can be regarded as a swarm of  A independent nucleons generating A superimposed proton-induced hadron cascades with energy $E/A$. Owing to the heavier CR nuclei possesses less energy for each nucleon, it can interact with the air molecules at higher altitude. Hence their shower electromagnetic components will suffer more attenuation with longer interaction length, and the $\pi^{\pm}$ components will have more opportunity to decay into the muon parts. As a consequence, the ratio of the electromagnetic to muon parts is a component-sensitive estimator and is adopted widely on the CR experiments \cite{kascade_heavy}.
\par
\begin{figure}
\includegraphics[width=.45\textwidth]{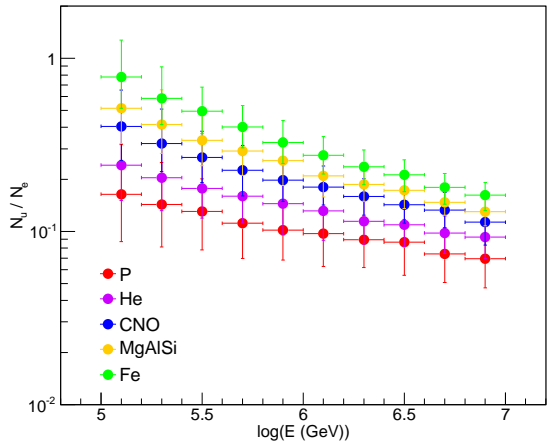}
\caption{\label{fig:nenu} The distribution of the value $N_{\mu} / N_e$ for each CR group (P (red), He (violet), CNO (blue), MgAlSi (yellow), Fe (green)) across the energy from 100 TeV to 10 PeV.}
\end{figure}

As the LHAASO-KM2A array can discriminate the electron and muon parts in the shower by the ED and MD arrays, we formulate the ratio of collected signals from the MD and ED $N_{\mu} / N_e$ as the physics-based baseline model. $N_{\mu}$ and $N_e$ denote the collected photoelectrons of an event recorded by the activated  MD and ED detectors respectively. The active ED detectors are counted within the distance 100 m from the shower core, while the active MD detectors are counted within the distance 40 $\sim$ 200 m from the shower core. The occlusion area within 40 m of the MD is to eliminating the punch-through effect, where the high-energy electronic particles near the shower core can penetrate the soil-shielding layer and fire the beneath MD detectors. We illustrate the distribution of the ratio $N_{\mu} / N_e$ with respect to CR components and energies in Fig. \ref{fig:nenu}. As shown, the heavier components possess  larger value of $N_{\mu} / N_e$, while Proton lies at the bottom.

\section{Graph Neural Network}
\subsection{Graph Neural Network Overview}
The GNN architectures are specialized to effectively analyze the graph-structured data, many of them borrow the idea from the convolution networks and design their graph convolution operations. Comparing with different graph convolution schemes, the major GNN models can be classified into two categories, including the spectral domain and the spatial domain \cite{gnn_overview}. The spectral methods are formulated based on the graph signal processing theory \cite{gsp, signal_graph_theory}, where the graph convolution is interpreted as filtering the graph signal on a set of weighted Fourier basis. The spatial methods explicitly aggregating the information from the neighbor through the weighted edges.
\par
In order to compare the differences between the spectral-domain and spatial-domain methods, we introduce the spectral graph processing theory at first. An undirected, connected, weighted graph is denoted as $G = \{ V, E, A\}$, which consists a set of vertices $V$, a set of edges $E$, and a weighted adjacency matrix $A$. The normalized graph Laplacian is defined as $L = I - D^{-1/2}AD^{-1/2}$, where the degree matrix $D$ is a diagonal matrix whose $i$th diagonal element $d_i$ is equal to the sum of weights of all the edges incident to vertex $i$. As the normalized graph Laplacian $L$ is a real symmetric positive semidefinite matrix, it can be factored as $L = U \Lambda U^T$, where $U = [ u_0, u_1, ..., u_{n-1}] \in R^{N \times N}$ is the matrix of eigenvectors ordered by the eigenvalues and $\Lambda$ is the diagonal matrix of eigenvalues, $\Lambda_{ii} = \lambda_i$. Mathematically, the eigenvectors forms an orthonomal space, where $U^T U = I$. Hence, the graph Fourier transform of a signal $x \in R^N$ is defined as $F(x) = U^T x$ and the inverse graph Fourier transform is defined as $F^{-1}(x) = U x$. The spectral-based method regards the graph convolution as the filtering operation on a set of Fourier basis of $x$, which has the form as
\begin{equation}
x * G \ g = U((U^T x) \odot (U^T g))
\label{eq:spectral_GNN_1}
\end{equation}
where $\odot$ is the element-wise Hadamard product.Denote a filter as $g_{\theta} = diag(U^T g)$, the convolution operation in Eq. \ref{eq:spectral_GNN_1} is simplified as
\begin{equation}
x * G \ g_{\theta} = U g_{\theta} U^T x
\label{eq:spectral_GNN_2}
\end{equation}
\par
All the spectral-base graph convolution operations follow the definition in Eq. \ref{eq:spectral_GNN_2}, except for the choice of $g_{\theta}$. Bruna et al. \cite{bruna_gnn} propose the first spectral convolution neural network (Spectral CNN), with the spectral filter $g_{\theta} = \Theta^k_{i,j}$ as a set of learnable parameters. As the computation complexity of the Fourier basis U is high $O(n^2)$, Defferrard et al. \cite{defferrard_gnn} propose the Chebyshev Spectral CNN (ChebNet) by introducing the Chebyshov polinomials as the filter, i.e. $g_{\theta} = \sum^K_{i=1} \theta_i T_k(\tilde{\Lambda})$, where $\tilde{\Lambda} = 2\Lambda / \lambda_{max} - I_N$. As a consequence, the ChebNet can avoid the computation of the graph Fourier basis, reducing the computation complexity to $O(K |\varepsilon|)$. Furthermore, Kipf et al. \cite{kipf_gnn} simplified the ChebNet as a first-order approximation by assuming the $K = 1$ and $\lambda_{max} = 2$. And the resulting graph convolution turns to be located totally into the spatial domain.
\par
The spatial-based graph convolution is defined based on the node's spatial relations. Following the idea of "correlation with template", the graph convolution relies on employing the local system at each node to extract the patches. Masci et al. \cite{gcnn} introduce the Geodesic CNN (GCNN) framework, which generalize the CNN into the non-Euclidean manifolds by mapping it into the local geodesic polar coordinate ($\rho$, $\theta$) around the position $x$. Boscaini et al. \cite{acnn} considered it as the anisotropic diffusion process, where the heat flow among the nodes is position- and direction-dependent.
\par
Generalizing many of the spatial-domain networks, Monti et al. \cite{monet} proposed the Mixture Model Networks (MoNet), a generic framework deep learning on the non-Euclidean domains. In this framework, a spatial convolution layer is given by a template-matching procedure as
\begin{equation}
(f * g)(x) = \sum^{J}_{j=1} g_j D_j(x) f
\label{eq:spatial_GNN}
\end{equation}
\par
And the patch operator in Eq. \ref{eq:spatial_GNN} take the form as
\begin{equation}
D_j(x)f = \sum_{y \in N(x)} \omega_j(u(x,y))f(y), j = 1, ..., J
\end{equation}
where J represent the dimension of the extracted graph. The $x$ denotes a point in the graph or the manifold, and $y \in N(x)$ represents the neighbors of the $x$. The $u(x,y)$ associates the node with the pseudo coordinate, and $\omega_j(u)$ is the weighting function parameterized by the learnable parameters.
\par
The definition of the patch operator associates the MoNet with other spatial-based graph convolutional models through the choice of the pseudo coordinate $u(x,y)$ and the weighting function $\omega_j(u(x,y))$. As a consequence, those spatial-based methods can be considered as the particular instances of the MoNet. In particular, a convenient choice of the weighting function is the Gaussian kernel
\begin{equation}
\omega_j(u) = \exp(-\frac12 (u - \mu_j)^T \Sigma^{-1}_j (u - \mu_j))
\end{equation}
where $\Sigma_j$ and $\mu_j$ are learnable $d \times d$ and $d \times 1$ covariance matrix and mean vector of a Gaussian kernel respectively.
\par
As can be seen, the spectral-based methods owns its mathematical foundation on the graph signal processing, but costs high computation in calculating the Fourier transform. The spatial-based methods is intuitive by directly aggregating the information from the neighbors, and has the potential to handle large graphs. On the other hand, owing to the Laplacian-based representation is required for the spectral convolution, a learned model is unable to be applied on another different graph, while the spatial-based convolution is flexible to be shared across different locations and structures. As the CR EAS event varies its location, direction and energy, the spatial-domain method is suitable in analyzing the LHAASO-KM2A experiment.

\subsection{Graph Neural Network On the LHAASO-KM2A}
The LHAASO-KM2A detectors can record the arrival time and the photoelectron amplitude of the shower secondary particles.  The distribution of detector photonelectrons with respect to the distance from the shower core roughly obeys the NKG function \cite{greisen_nkg, nk_nkg} with the most dense region located at the shower core, while the distribution of arrival times can be parameterized as a plane perpendicular to the direction of the shower. Accordingly, we can make the data preprocessing procedure by reconstructing the event to locate the shower core position ($x_0$, $y_0$) and the direction ($\theta_0$, $\phi_0$). The photoelectrons is normalized to the reconstructed event energy for the energy-invariant representation, denoted as $pe$. As the shower geometry is often treated as a slanted symmetric plane around the shower core,  we transform the detector positions ($x_i$, $y_i$) into the cylinder coordinate ($r_i$, $\phi_i$) with the zero point at the shower core. The shower event along the time axis is represented by the detector's time residual $dT_i$ defined as
\begin{equation}
dT_i = T_i  - \frac{{\bf r_i} \cdot {\bf r_0}}{c  \lVert {\bf r_0} \rVert} - T_0
\end{equation}
where $T_i$ is the recording time by the detector, and $T_0$ is the reference time., $r_i$ and $r_0$ represent the vectors of the node position and shower direction respectively, and $c$ is the speed of light.
\begin{figure}
\includegraphics[width=.45\textwidth]{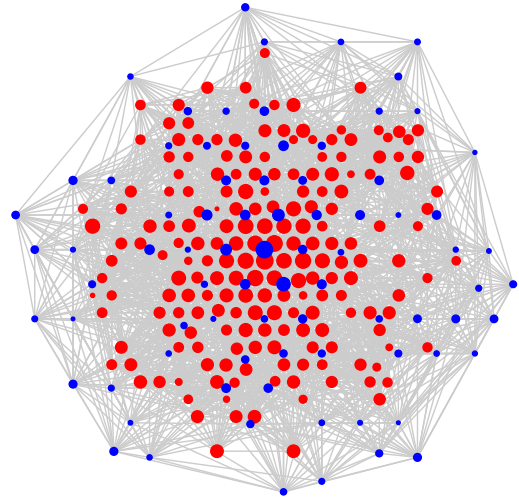}
\caption{The graph-structured LHAASO-KM2A detectors activated by a 500-TeV EAS event, where the red dots represent EDs and the blue dots represent MDs. The dot size measures the logarithmic scale of the recorded photoelectrons.}
\label{fig:event}
\end{figure}
\par
The ED and MD detectors are constructed as weighted, undirected dense graphs independently, with each node contains a 3-dimension vector $[pe_i$, $dT_i$, $r_i]$. This 3-dimension feature is normalized for each channel independently. An event graph is shown in Fig. \ref{fig:event} for illustration. We construct the GNN model similar to the works in \cite{icecube_gnn, monet}. The $n \ \times \ n$ adjacency matrix A is defined by applying the Gaussian kernel to the pairwise distance $\lVert x_i - x_j \rVert$ between the activated detectors as follows
\begin{align}
& d_{ij} = e^{- \frac12 (\lVert x_i - x_j \rVert - \mu_t)^2 / \sigma_t^2} \\
& a_{ij} = \frac{d_{ij}}{\sum\limits_{k \in N} d_{ik}}
\label{eq:weight}
\end{align}
\par
In Eq. \ref{eq:weight}, $a_{ij}$ is the normalized weight element in the adjacency matrix, and $N$ represents the set of adjacent detectors with respect to the detector $i$. The $\mu_t$ and $\sigma_t$ are learnable parameters, which defines the locality of the convolutional kernel. In addition, the diagonal elements in the matrix A is set to zero.
\par
Before implementing the graph convolution layers, we extract the higher-dimensional features from the input vectors through the learnable function as shown in Eq. \ref{eq:layer_feature}, from which the $n \ \times \ 3$ vertex matrix $v$ converts into the $n \ \times \ d^{(0)}$ matrix $x^{(0)}$.
\begin{equation}
x^{(0)} = ReLu(W^{(0)} v + b^{(0)})
\label{eq:layer_feature}
\end{equation}
\par
Then, we defined a sequence of T convolution layers following as shown in Eq. \ref{eq:conv}. Each convolution layer $t$ firstly aggregates the neighbors by the multiplication with the adjacency matrix A and expands the vector from $d^{(t)}$- to $2 d^{(t)}$- dimension. Next, the weighting function is applied to update the vector into $d^{(t+1)}$- dimension. The nonlinear activation function $ReLu$ is used, expect for the last convolution layer $T$.
\begin{align}
& GConv(x^{(t)}) = W^{(t)} [ x^{(t)}, A x^{(t)} ] + b^{(t)} \\
& {x^{(t+1)} = \begin{cases}
ReLu(GConv(x^{(t)})), & t+1 < T \\
GConv(x^{(t)}), & t+1 = T
\end{cases}}
\label{eq:conv}
\end{align}
\par
The graph structure is preserved during the convolutional operations. At the last convolution layer, i.e. the $T$th layer, we add a global pooling layer to collect the feature across the whole graph and compress the graph into a size-invariant representation. The $n \times  d^{(T)}$ feature matrix is averaged and converted into $1 \times d^{(T)}$-dimensional matrix. The definition of the global pooling layer is
\begin{equation}
x_i^{(pool)} = \frac1N \sum\limits_{n \in N} x_{ni}^{(T)}
\end{equation}
At the last layer, we use a linear layer and the logistic regression is applied to evaluate the event score as the classifier,
\begin{equation}
y = sigmoid( W^{(pool)} x^{(pool)} + b^{(pool)} )
\label{eq:score}
\end{equation}
where $x^{(pool)}$ is the $d^{(T)}$-dimensional feature from the global pooling layer and $y$ is the voting score. The activation function $sigmoid$ ensures the score y spreads within the range $[0,1]$, where the signal-like or background-like event approaches 1 or 0 respectively.
\par
\begin{figure*}
\centering
\includegraphics[width=.8\textwidth]{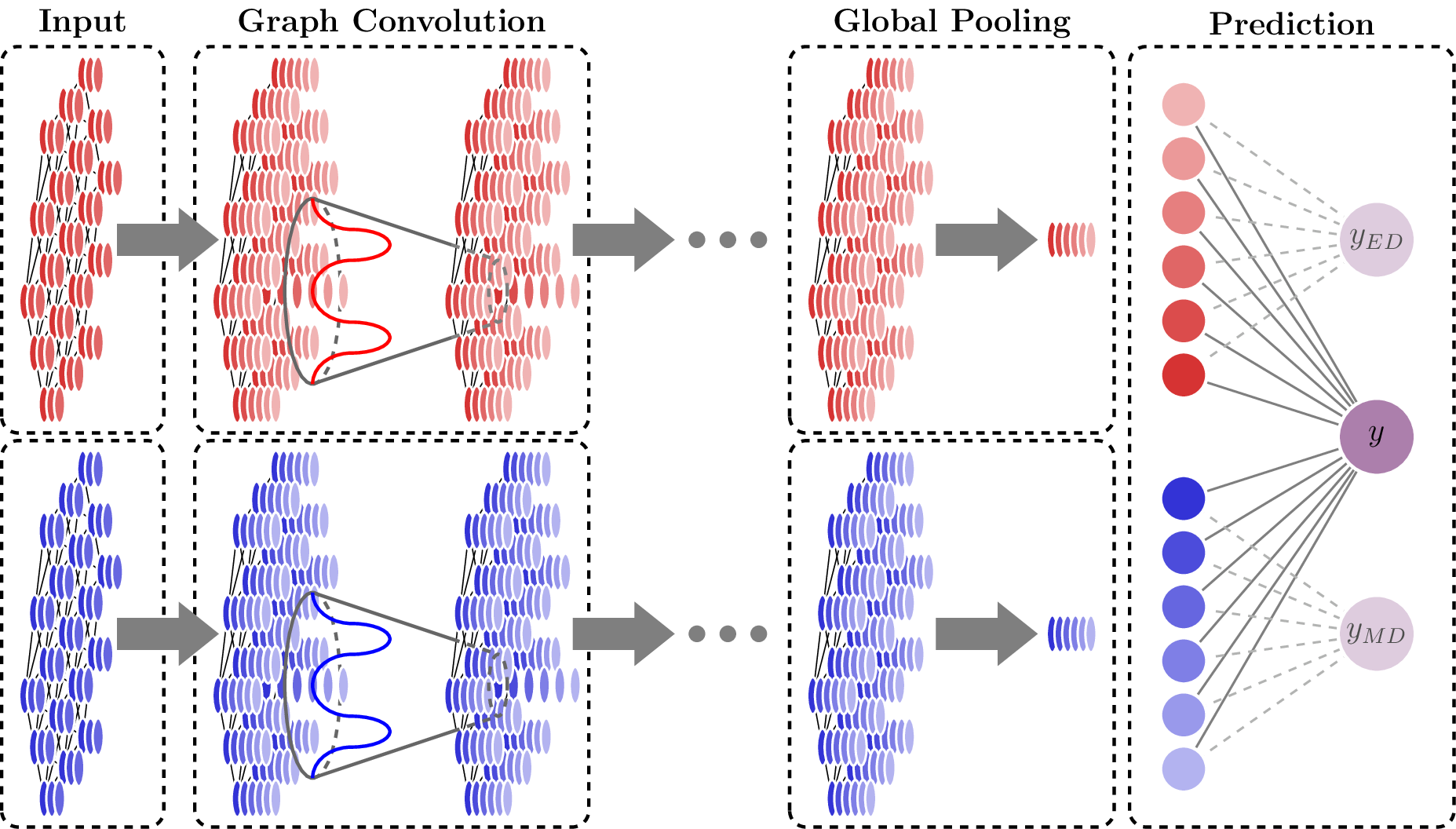}
\caption{\label{fig:gnn} The KM2A GNN model with the upper red network represents the GNN ED model, and the lower blue network represents the GNN MD model. The right most rectangular contains the fusion operation of the two models (GNN ED+MD) and their independent outputs.}
\end{figure*}

We construct the GNNs for ED and MD independently, and fuse their outputs together through the linear layer in Eq. \ref{eq:score} with the $x^{(pool)}$ is a $2d^{(T)}$-dimensional vector. Independent GNN models for the ED and MD are preserved for comparison. The total GNN architecture is illustrated in Fig. \ref{fig:gnn}.

\section{Experiment}
In this work, we use the Monte Carlo simulation to generate the event data for training and evaluating the KM2A GNN performance. The primary EAS events are generated by the CORSIKA package \cite{corsika}. And the KM2A detector simulation is performed based on the Geant4 framework \cite{km2a_geant4}. We generate the major CR groups including the Proton (P), the Helium (He), the medium group (CNO), the heavy group (MgAlSi), and the Iron (Fe). Total events are generated into 4 energy fragments, including 10 TeV $\sim$ 100 TeV, 100 TeV $\sim$ 1 PeV, 1 PeV $\sim$ 10 PeV, 10 PeV $\sim$ 100 PeV, with the spectral index  -2.7. And the reconstructed energies from 100 TeV to 10 PeV are considered, which cover the major CR knee region. As for each task, these groups are divided into the signal and background groups independently, where only P belongs to the signal for the P task, and P\&He forms the signal for the L task.
\par
After the reconstruction for the simulated events, we further select the events according to their reconstructed locations and directions. The reconstructed shower core spreads inside the KM2A array within the distance 200 $\sim$ 500 m from the array center are selected. We ignore the inner circular area (within 200 m) in order to suppress the disturbance from the WCDA for the KM2A reconstruction. In addition, the reconstructed zenith angle below $35^{\circ}$ is also required. As a consequence, it remains 105732 events for the next analyzing. We split the selected events into the train, test and evaluation data sets. In consideration for the data balance, the group ratios for each data set are readjusted to maintain the $1:1$ signal-to-noise ratio (SNR). The readjusted data sets for each task are listed in Tab. \ref{tab:dataset}. And the dataset ratio between the two major energy fragments, with 100 TeV $\sim$ 1 PeV and 1 PeV $\sim$ 10 PeV, is around $2:1$.
\par
To train the GNN models, we use the supervised learning techniques with the mean square error (MSE) as the loss function. The Adam \cite{adam} optimizer is used to optimize the model parameters based on the adaptive estimation of the low-order moments. The training procedure include two steps, (i) two independent trainings for the GNN ED and MD models with the learning rate 0.001 (ii) and a following fine tuning procedure fuses the ED and MD model together with the learning rate 0.0001. It runs over 80 epoch totally with the model already converged. All these code are written in Python using the open-source deep learning framework PyTorch with the GPU acceleration. For each model training, 4 same candidates with different randomized weights are trained and the one with the best performance is selected for the further process.

\begin{table}
\centering
\caption{\label{tab:dataset} The number of the signal and background events for each data set}
\begin{tabular}{ccccc}
  \toprule
  \hline
  \multirow{2}{*}{data set} & \multicolumn{2}{c}{P} & \multicolumn{2}{c}{L}\cr\cline{2-5}
                         & Signal & Background & Signal & Background \cr
  \hline
  Train         & 14635 & 14595 & 24358 & 23733 \\
  Test          & 2875  & 2831  & 4754  & 4713  \\
  Evaluation    & 24921 & 22994 & 24921 & 22994 \\
  \hline
  \bottomrule
\end{tabular}
\end{table}

\section{Results}
\begin{figure*}
\includegraphics[width=.45\textwidth]{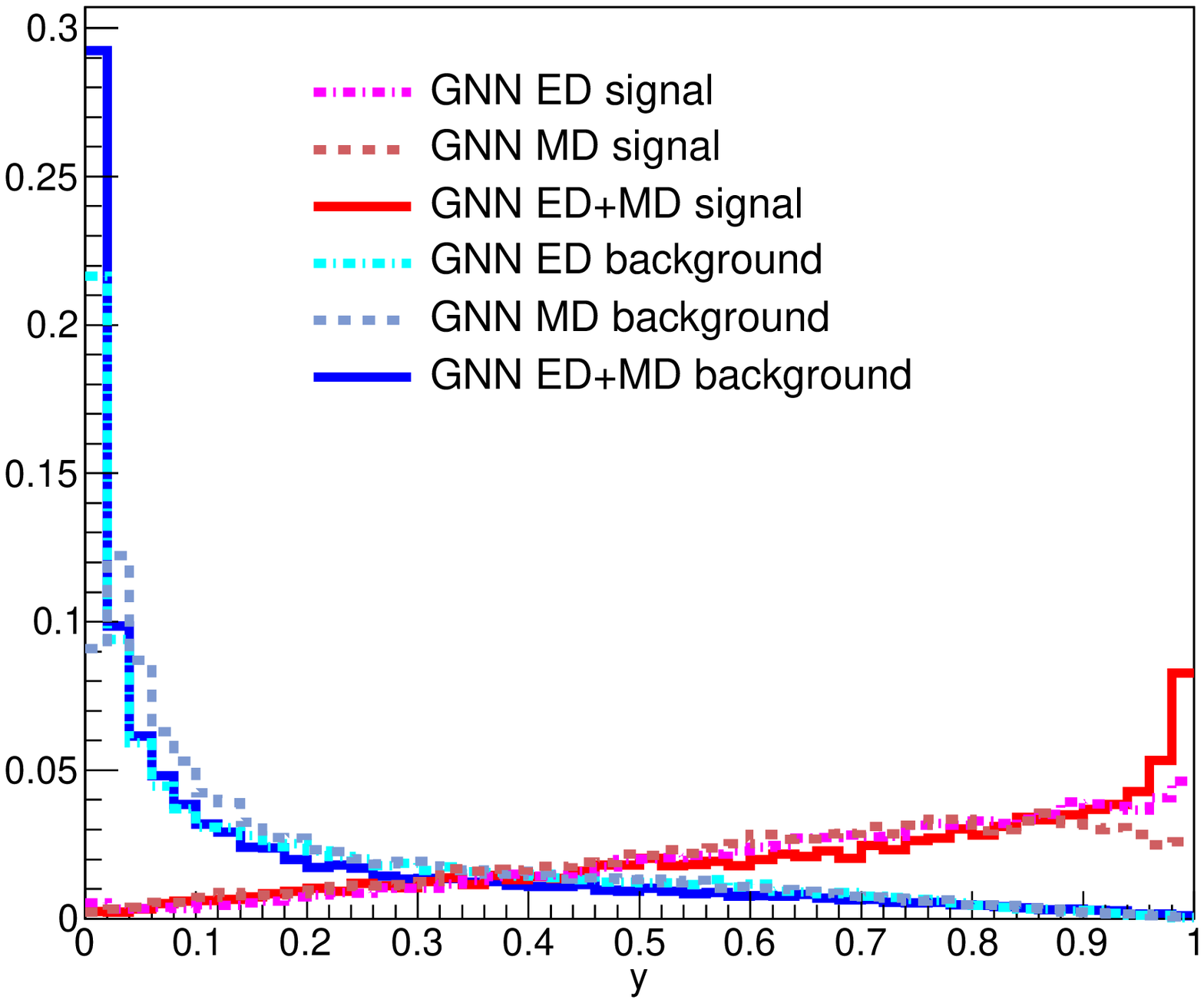}
\includegraphics[width=.45\textwidth]{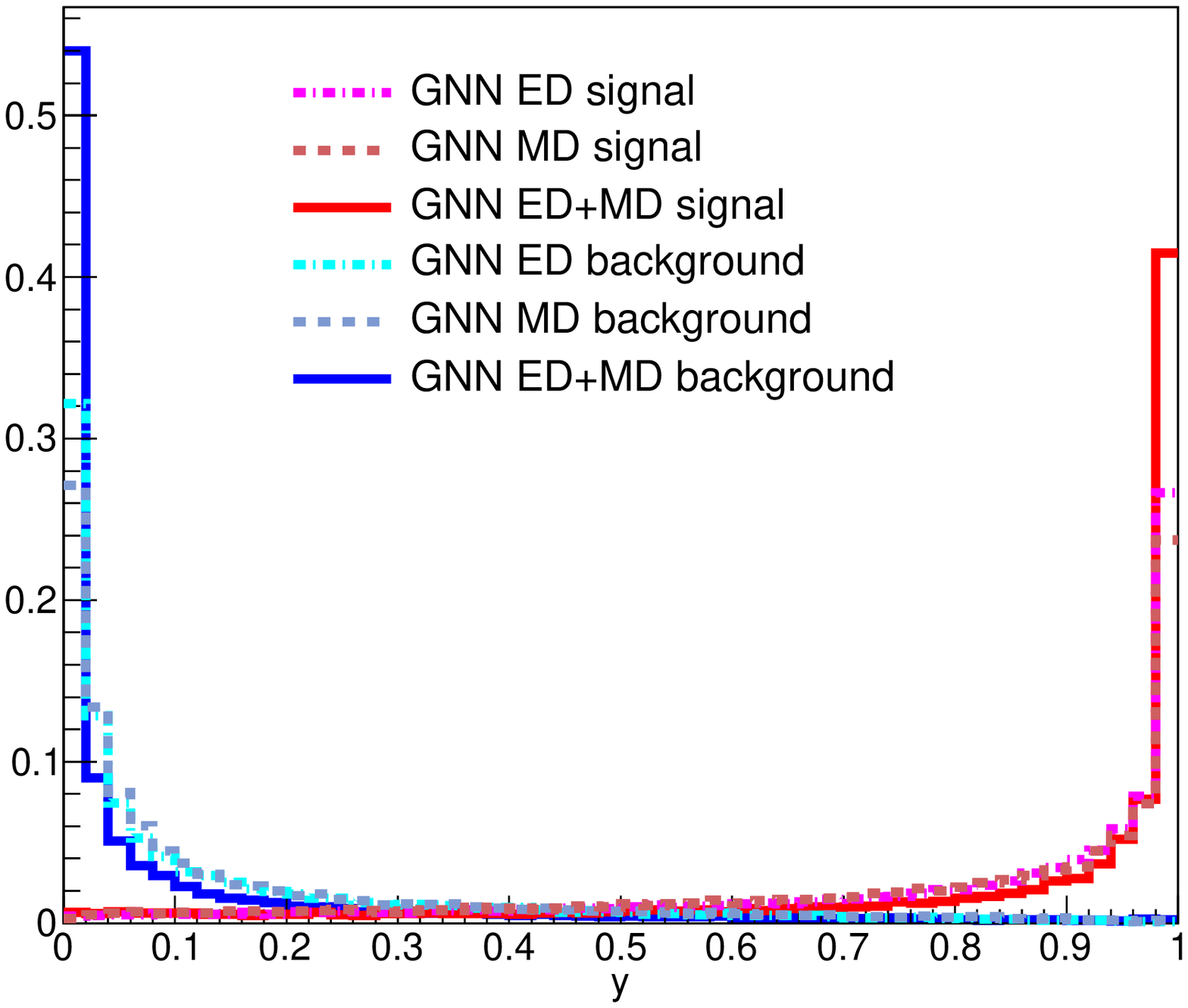}
\caption{The distribution of the output scores from each model for the P task (left) and L task (right)}
\label{fig:score}
\end{figure*}

\begin{figure*}
\includegraphics[width=.45\textwidth]{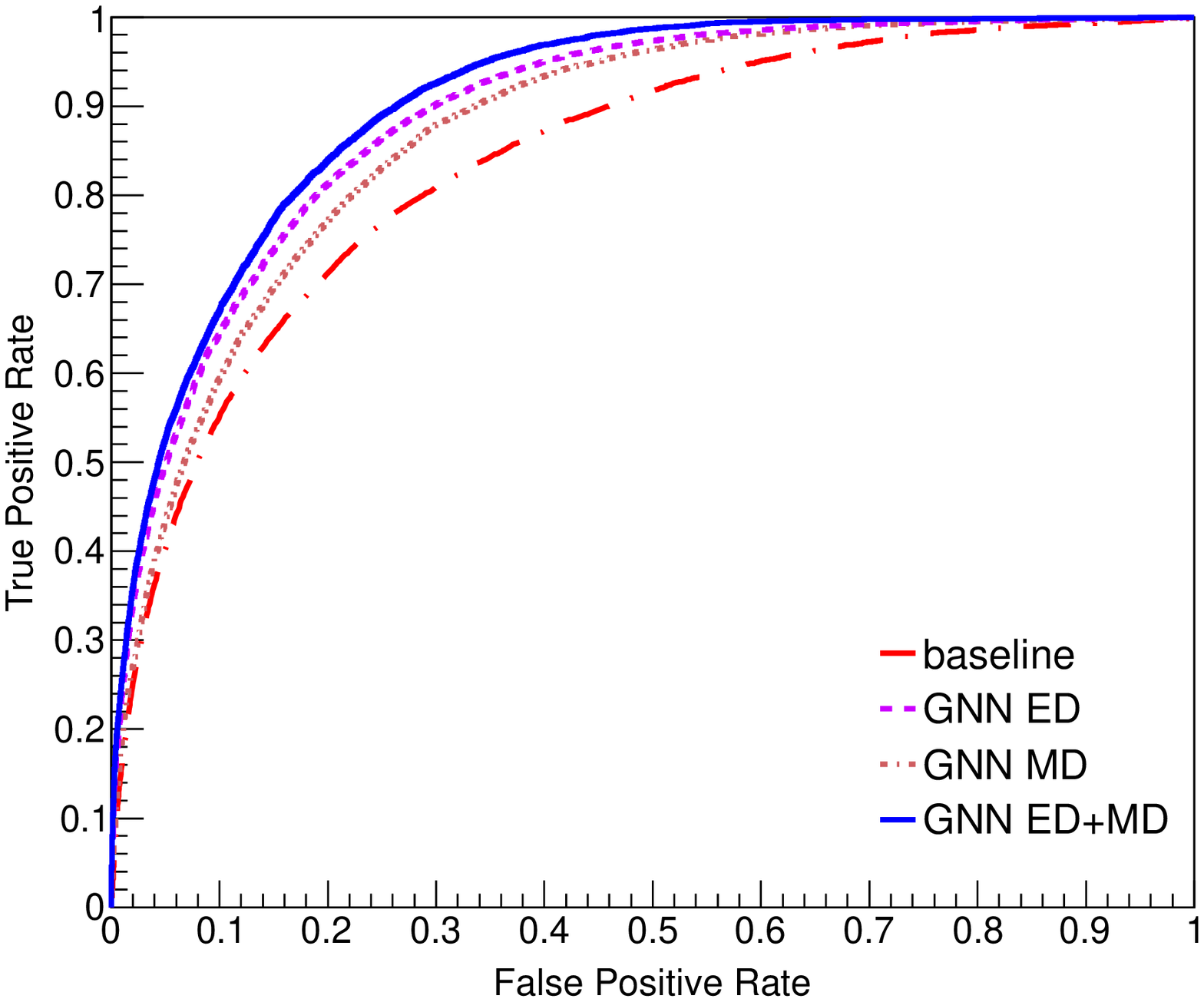}
\includegraphics[width=.45\textwidth]{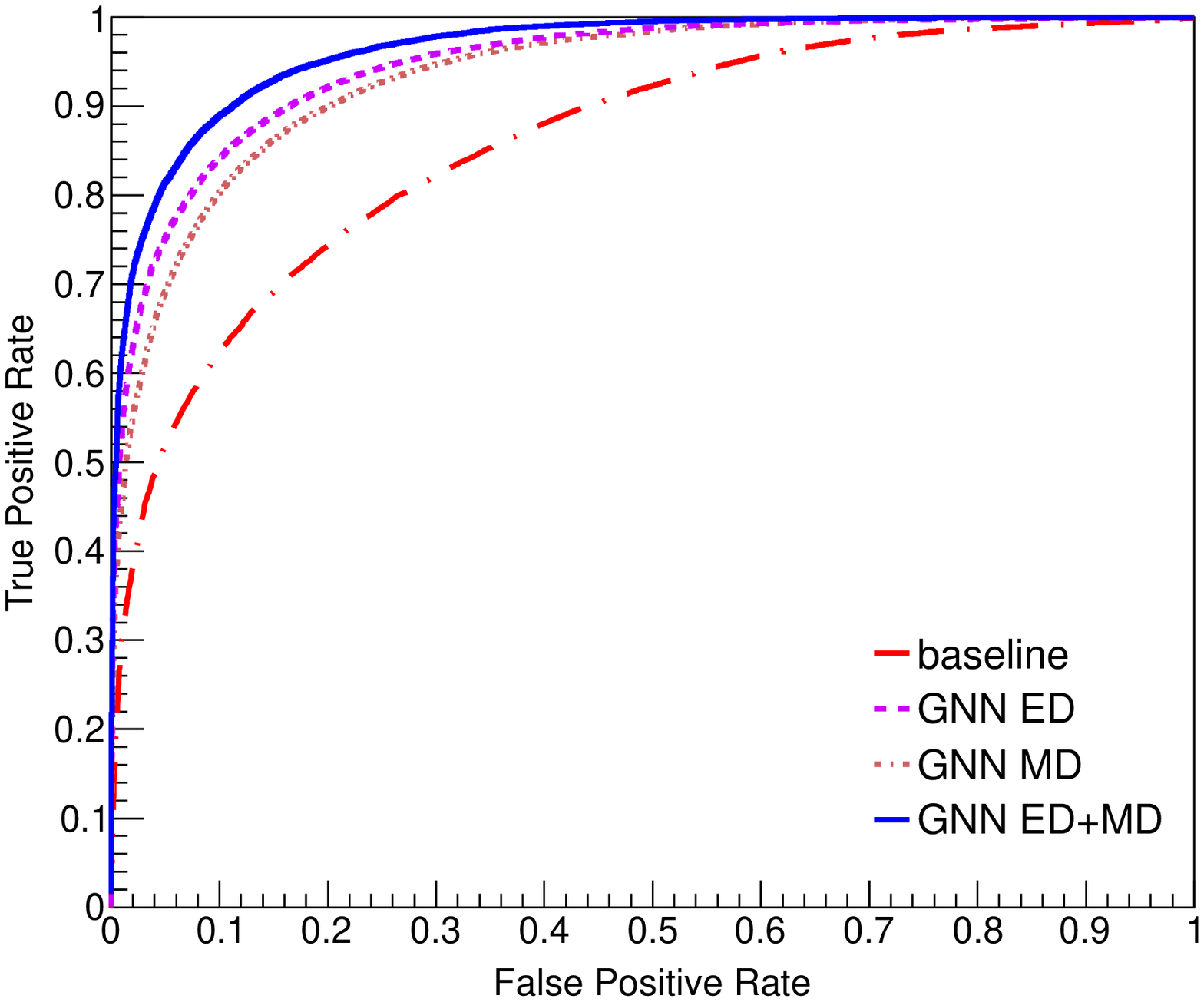}
\caption{The ROC curves from each model for the P task (left) and L task (right)}
\label{fig:roc}
\end{figure*}

We evaluate the model performances on the evaluation dataset for each task. The Fig. \ref{fig:score} displays the distributions of the output scores. All results from the P and L tasks are depicted at the left and right channels respectively. Intuitively, the shapes of the score indicate the task for classifying the light group is much easier than the singular proton group. We calculate the receiver operating characteristic (ROC) curves for the explicit comparison. The ROC curves of the physics baseline are integrated on the $N_{\mu}/N_e$ distribution, while the curves of the GNN models are integrated on the scores. Results are shown in Fig. \ref{fig:roc}. The ROC's x axis, called the False Positive Rate, means the  background retention rate. And the ROC's y axis, called the True Positive Rate, means the signal efficiency. The Fig. \ref{fig:roc} clarifies that the best performance is the fused GNN model, while the physics baseline performs the worst performance. And the ED GNN model behaves better than the MD GNN, from which we consider the sparsity of the sub-array takes the essential effect.

\par
\begin{figure*}
\includegraphics[width=.45\textwidth]{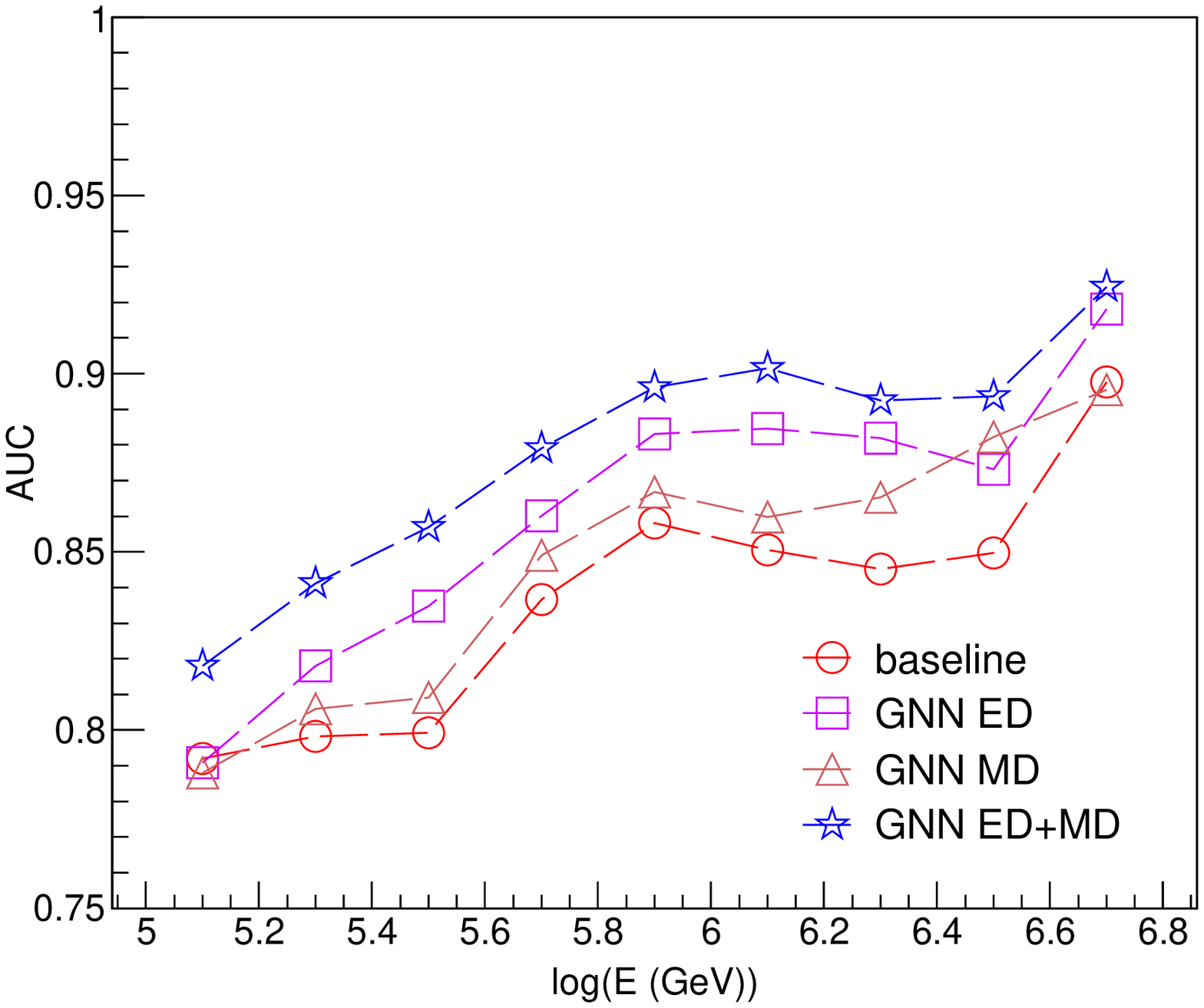}
\includegraphics[width=.45\textwidth]{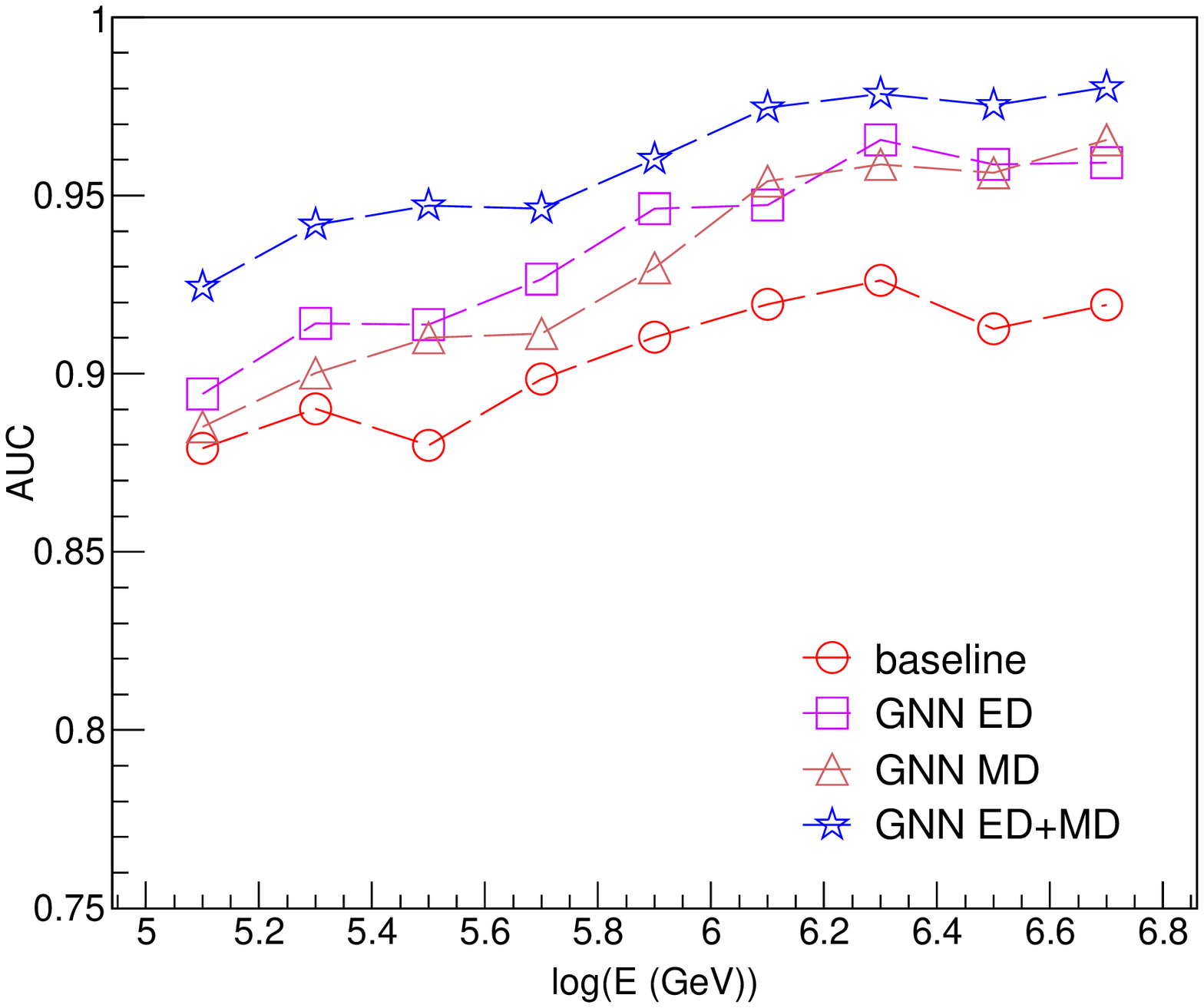}
\caption{The AUC values across the energy range from 100 TeV to 10 PeV from each model for the P task (left) and L task (right)}
\label{fig:auc}
\end{figure*}

In order to reduce the sensitivity to noise due the limited volume of data set, as well as evaluate the model performance quantitatively, we use the area under the ROC (AUC) curves as a measure of the model performance. We further split the data set into a sequence of energy bins for comparison of the performance across the whole energy range. For the energy range with one order of magnitude, we split it into 5 bins uniformly in the logarithmic coordinate. We calculate the AUC values of the models at each bin with each of the selected event weighted by the Horandel model \cite{horandel}, and plotted them in the Fig. \ref{fig:auc}. As shown, the results confirm those conclusions announced above. Furthermore, it also manifests that the fused GNN model outperforms the physics baseline across the whole energies. We average the AUCs values, and list them in Tab. \ref{tab:auc}. The fused GNN model achieve the highest score with 0.878 for the P task, and 0.959 for the L task. It can be seen that the AUC score of the L task always exceed the P task by a considerable amount of value, which is about 0.068 in the physics baseline and rises up to 0.081 in the fused GNN model. As the nuclear number of the Proton and Helium are close, it is hard to discriminate the Proton from the Helium background.


\begin{figure}
\includegraphics[width=.45\textwidth]{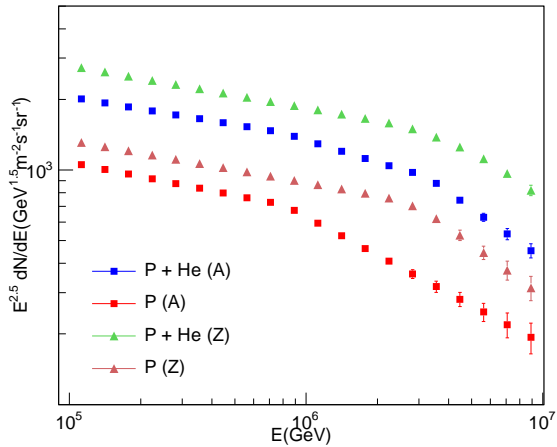}
\caption{Expectation on the proton- and light-group spectra measured by the LHAASO-KM2A with 1-day observation. The triangular markers represent the spectra predicted by one of the rigidity-dependent knee models (Z) \cite{gamma_knee}, and the square markers represent the spectra predicted by one of the mass-dependent knee models (A) \cite{lhaaso_knee}}
\label{fig:knee}
\end{figure}

\begin{table}
\centering
\caption{\label{tab:auc} The average AUC scores}
\begin{tabular}{ccc}
  \toprule
  \hline
   & P & L \\
  \hline
  baseline        & 0.836       & 0.904       \\
  GNN MD          & 0.847       & 0.93        \\
  GNN ED          & 0.861       & 0.936       \\
  {\bf GNN ED+MD} & {\bf 0.878} & {\bf 0.959} \\
  \hline
  \bottomrule
\end{tabular}
\end{table}
\par
At the same time, strategies based on the LHAASO hybrid detection have been explored, which explore the latent representation in combination all of the WCDA, KM2A and WFCTA arrays. Those strategies include making a handcrafted selection criteria \cite{lhaaso_hand}, and the traditional machine learning method such as the Gradient Boosted Decision Tree (GBDT) \cite{lhaaso_gbdt}. In comparison with the hybrid method, we set the models in this work at the roughly same condition and compare their signal purity and the array apertures. Those values are listed in Tab. \ref{tab:comp}. It can be seen that the purity implicitly indicates the KM2A GNN performance is competitive to the hybrid method. The aperture relies on the array configuration, with the WFCTA's field of view ($15^{\circ} \times 15^{\circ}$), the smaller area of the WCDA (78000 $m^2$), and the large area of the KM2A (1.3 $km^2$). As shown in Tab. \ref{tab:comp}, the aperture of the KM2A can acquire the order of 87$\times$ larger than the hybrid detection. Considering the WFCTA's strict observation condition with the $\sim$ 10\% duty cycle \cite{lhaaso_hand}, the total statistics of the KM2A is expected to be the order of 870$\times$ larger than the hybrid detection. We illustrate the expected observation with 1-day operation of the KM2A on the proton- and light-group spectra in Fig. \ref{fig:knee}, where the rigidity- and mass-dependent knee models are adopted from \cite{lhaaso_knee, gamma_knee}.
\par
We further construct a simple CNN model for comparing with the GNN model. The whole ED and MD arrays are rescaled into regular grids respectively, with $(85 \times 97)$ pixels for ED and $(40 \times 46)$ pixels for MD. And activated detectors are filled into corresponding grids with others remain zero. We construct the CNN model with series of convolution modules for the ED and MD images, and fuse their output together through a linear layer as the classifier. The performance is shown in Tab. \ref{tab:comp} as well. It demonstrates that the CNN exhibit a poor performance, from which we attribute it to the insufficient ability in analyzing large variance of the EAS (10s $\sim$ 1000 activated detectors) and inefficient representation of the image structure (zero grids $\gtrsim$ 90\%). On the other hand, as the graph convolutional kernel in Eq. \ref{eq:weight} is the Gaussian function with only 2 learnable parameters ($\mu_t$, $\sigma_t$), while the number of parameters in a CNN convolutional layer is $C_{out} \times C_{in} \times K_t^2$, the training efficiency of the CNN is far less than the GNN.

\begin{table}
\centering
\caption{\label{tab:comp} The signal purity and the aperture values of each model on the LHAASO experiment}
\begin{tabular}{ccccc}
  \toprule
  \hline
  \multirow{2}{*}{} & \multicolumn{2}{c}{Purity (\%)} & \multicolumn{2}{c}{Aperture ($m^2 \cdot sr$)}\cr\cline{2-5}
                         & P & L & P & L \cr
  \hline
  handcraft (hybrid) \cite{lhaaso_hand}         & 90 & 95 & 1.5e3 & 4e3 \\
  GBDT (hybrid) \cite{lhaaso_gbdt}         & 90  & 97  & 3.6e3  & 7.2e3  \\
  baseline (KM2A)    & 73.4 & 93.2 & 3.2e5 & 6.3e5 \\
  CNN (KM2A)       & 75.4 & 93.3 & 3.2e5 & 6.3e5 \\
  GNN MD (KM2A)   & 77.1 & 95.9 & 3.2e5 & 6.3e5 \\
  GNN ED (KM2A)    & 82.8 & 96.6 & 3.2e5 & 6.3e5 \\
  {\bf GNN ED+MD (KM2A)}    & {\bf 84} & {\bf 98.2} & {\bf 3.2e5} & {\bf 6.3e5} \\
  \hline
  \bottomrule
\end{tabular}
\end{table}

\section{Conclusion}
Realizing that a great deal of progresses have been achieved on the deep learning in many fields, we leverage this technology to improve the classification performance on the LHAASO -KM2A experiment. We propose a fused GNN model, which construct independent networks for the KM2A ED and MD arrays, and fuse their outputs for classification. It is manifested that this model is effective and the performance outperforms the traditional physics-based method as well as the CNN-based method across the whole energy range. In addition, we also compare the performance of the GNN framework for independent ED and MD array. It is found that ED array behaves better than the MD array. And we attribute this to the denser configuration of the ED array. Furthermore, comparing with the LHAASO hybrid detection method, our KM2A GNN model shows competitive classification performance. Benefit from the large area and full duty cycle of the KM2A array, it can acquire the statistics on the order of $\sim$ 870$\times$ larger than the hybrid detection.

\section{Acknowledgements}
We thank the LHAASO Collaboration for their support on this project. This work is supported by the National Key R\&D Program of China (number 2018YFA0404201), the Natural Sciences Foundation of China (numbers 11575203, 11635011).


\end{document}